\documentclass[aps,prl,reprint,superscriptaddress,longbibliography]{revtex4-2}
\usepackage{graphicx}
\usepackage{textcomp}
\usepackage{gensymb}
\usepackage{amsmath, amssymb}
\usepackage[colorlinks = true, citecolor = magenta]{hyperref}
\usepackage{braket}
\usepackage{natbib}
\usepackage[utf8]{inputenc}

\begin{document}

\title{STIRAP-inspired robust gates for a superconducting 
dual-rail qubit}

\author{Ujjawal~Singhal}
\author{Harsh~Vardhan~Upadhyay}
\author{Irshad~Ahmad}
\author{Vibhor~Singh}
\email{v.singh@iisc.ac.in} 
\affiliation{Department of Physics, Indian Institute of Science, Bangalore-560012 (India)}

\date{\today}

\begin{abstract}

STImulated Raman Adiabatic Passage (STIRAP) is a powerful 
technique for robust state transfer capabilities in quantum systems.
This method, however encounters challenges for its implementation 
as a gate in qubit-subspace due to its sensitivity to initial states. 
By incorporating single-photon detuning into the protocol, 
the sensitivity to the initial state can effectively be 
mitigated, enabling STIRAP to operate as a gate. 
In this study, we experimentally demonstrate the implementation of 
robust $\pi$ and $\pi/2$ rotations in a dual-rail qubit formed 
by two strongly coupled \textit{fixed-frequency} transmon qubits. 
We achieve state preparation fidelity in excess of 
0.98 using such rotations. Our analysis reveals these gates 
exhibit significant resilience to errors. 
Furthermore, our numerical calculations confirm that these gates 
can achieve fidelity levels in excess of 0.999. 
This work suggest a way for realizing quantum gates 
which are robust against minor drifts in pulse or system parameters.

\end{abstract}

\maketitle

\section{Introduction}

Accurate and efficient control over a quantum system
is essential for developing robust quantum computing platforms.
Typically, such control is achieved by driving the system with 
\textit{resonant} pulses and manipulating their parameters like amplitude, 
phase, and frequency. 
On one hand, while resonant interactions are accurate, 
they are also highly sensitive to parameter variations.
Adiabatic evolution techniques, on the other hand, offer 
greater robustness against minor drifts in 
both \textit{pulse} and \textit{system} parameters 
\cite{zanardi_holonomic_1999, duan_geometric_2001}.
STIRAP is a powerful 
technique among adiabatic protocols and it has 
been widely used for population transfer and 
precision measurements across 
various systems such as Rydberg atoms, 
trapped ions, and various solid-state systems 
including superconducting
qubits
\cite{vitanov_stimulated_2017, bergmann_roadmap_2019,kumar_stimulated_2016, xu_coherent_2016, 
vepsalainen_superadiabatic_2019, siewert_adiabatic_2006}. 
While STIRAP protocols are highly robust for population 
transfer and creating superposition states \cite{vitanov_creation_1999}, 
they do not directly constitute quantum gates due to 
their sensitivity to the initial state
\cite{vitanov_stimulated_2017, bergmann_roadmap_2019}.
This limitation can be overcome with modifications 
to the protocol. For instance, introducing a non-zero 
single-photon detuning and using 
fractional-STIRAP with back-to-back pulses can lead to robust swap gates 
\cite{lacour_arbitrary_2006, beterov_quantum_2013, genov_robust_2023}. 
Additionally, using auxiliary levels can facilitate the realization of 
rotations in the qubit subspace \cite{kis_qubit_2002, moller_geometric_2007}.
In the case of superconducting qubits, there is recent 
interest in developing STIRAP-inspired gates to connect 
states with low transition dipoles \cite{ribeiro_accelerated_2019, setiawan_analytic_2021}. 
Such protocols have also been investigated for implementing 
one- and two-qubit gates, contributing to the broader class of geometric quantum gates 
\cite{abdumalikov_jr_experimental_2013, xu_experimental_2020, chen_error-tolerant_2022, 
howard_implementing_2023, fang_nonadiabatic_2024, setiawan_fast_2023}.
However, STIRAP and its super-adiabatic variants have been 
experimentally utilized to demonstrate the population transfer 
in a single superconducting qubit \cite{kumar_stimulated_2016, xu_coherent_2016, 
vepsalainen_superadiabatic_2019, siewert_adiabatic_2006}. 
Thus far, their application 
as a quantum gate in superconducting platform has remain unexplored.

Recently, resonantly coupled transmons have been used to realize dual-rail 
qubits, where the computational subspace is spanned by the symmetric and 
anti-symmetric combinations of the single-excitation states 
\cite{kubica_erasure_2022, levine_demonstrating_2024}. 
Due to the low transition dipole of the two "logical" states, the dual-rail qubit exhibits higher 
coherence times, as the reduced dipole suppresses direct radiative transitions. 
It also makes it challenging to drive the qubit with a resonant pulse. 
Thus, a dual-rail qubit could be a suitable platform for testing
the performance of STIRAP-inspired quantum gates. 
So far, the single-qubit gates in dual-rail qubit have
been implemented by designing one of the qubits to be 
flux-tunable and modulating its frequency 
\cite{levine_demonstrating_2024,caldwell_parametrically_2018}.

In this work, we demonstrate STIRAP-inspired gates in a dual-rail 
qubit, marking the experimental realization 
of both $\pi$ and $\pi/2$ gates in superconducting qubits using 
detuned STIRAP.
While $\pi$ rotations using detuned STIRAP have been theoretically 
proposed, our study represents the experimental demonstration 
of such gates. 
Additionally, we extend this methodology to implement $\pi/2$ gates
using detuned STIRAP, which have not been explored earlier. 
By employing large single-photon detuning, we achieve robust 
gate operations. 
These gate operations are validated through quantum state 
tomography, and their performance is evaluated with respect 
to deviations in Rabi amplitude and frequency. 
It is interesting to remark that while the protocol operates 
within a single-qubit subspace, it is inherently a two-qubit 
operation in the individual qubit basis.
In our protocol, the use of the common ground state as an 
auxiliary level, and two microwave pulses can reduce the 
hardware requirements and minimize environmental noise
in a dual-rail qubit architecture. 

The paper is organized as follows: we first provide 
a brief overview of the STIRAP protocol 
to establish notation and to convey the basic idea. 
Next, we discuss the case of detuned STIRAP, 
showing how detuning in the control signals enables 
single-qubit rotations in a subspace 
spanned by a three-level system while maintaining partial adiabaticity. We then present 
experimental results, discussing the single qubit rotations on Bloch sphere and their robustness to control pulse parameters.

\section{Theoretical model}

\subsection{Brief overview of STIRAP}

\begin{figure}
\centering
\includegraphics[width=85mm]{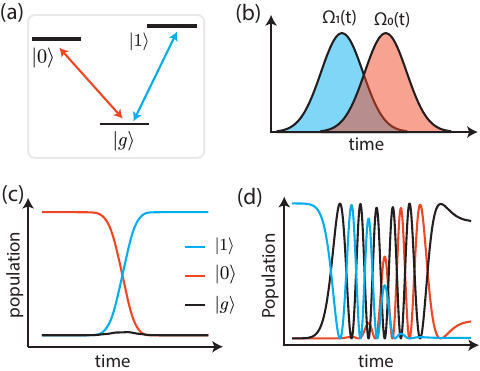}
\caption{\textbf{Schematic of resonant STIRAP} (a) A schematic of the V-type 3-level system
consisting of a ground state $\ket{g}$, and two excited states
$\ket{0}$ and $\ket{1}$. (b) A schematic of a counter-intuitive 
pulse ordering for the transfer of population from $\ket{0}$ 
to $\ket{1}$. 
(c) Time-evolution of the populations in different 
levels for the resonant STIRAP when the initial state is $\ket{0}$. 
Ultimately, it leads to a complete transfer of population 
to $\ket{1}$. (d) The evolution when the initial 
state is $\ket{1}$. The oscillations show the mixing between
the three states as the system evolves.}
\label{fig:fig1_stirap_concept}
\end{figure}

STIRAP is a control technique allowing complete 
population transfer between two uncoupled states 
via one or more intermediate states that are never 
populated ideally. 
Consider a general ``V-type" system comprising of 
two orthogonal excited states, denoted by $\ket{0}$ and 
$\ket{1}$, and a ground state denoted as $\ket{g}$ as 
shown in Fig.~\ref{fig:fig1_stirap_concept}(a).
The transitions $\ket{g} \leftrightarrow \ket{0}$ 
and $\ket{g} \leftrightarrow \ket{1}$ can be driven 
by two-time dependent signals ${\Omega}_0(t)\cos\omega_{0}t$ 
and ${\Omega}_1(t)\cos\omega_{1}t$. 
The control signals are characterized by slowly 
varying Rabi amplitudes $\Omega_0(t)$ and $\Omega_1(t)$ 
as shown in Fig.~\ref{fig:fig1_stirap_concept}(b). 
Here, $\omega_0$ and $\omega_1$ are the transition 
frequencies corresponding to $\ket{g} \leftrightarrow \ket{0}$ 
and $\ket{g} \leftrightarrow \ket{1}$ transitions, respectively.

By performing the rotating wave transformation to eliminate
the fast-time dependence on the control fields, it 
is straightforward to obtain the following Hamiltonian
with \(\hbar = 1 \),
\begin{align}
H(t) = \frac{1}{2}\begin{pmatrix} 0 & \Omega_{0}(t) & 0 \\
\Omega_{0}(t) & 2\Delta & \Omega_{1}(t) \\
0 & \Omega_{1}(t) & 2\delta \\
\end{pmatrix},
\label{eq:H0}
\end{align}
where the detunings of the driving fields from the 
corresponding resonances are defined as 
$\Delta_p = \omega_{g0} -\omega_{0}$ and 
$\Delta_s = \omega_{g1} -\omega_{1}$ 
with the single-photon detuning given by 
$\Delta = (\Delta_p + \Delta_s)/2$, and the two-photon detuning 
$\delta = \Delta_p - \Delta_s$. 
Insight into the dynamics of the system can be made by
setting the two-photon detuning $\delta$ = 0. 
This leads to the following instantaneous eigenstate, 
\begin{align*}
\ket{+} &= \sin \theta \sin \phi \ket{0} + \cos \theta \sin \phi \ket{1} + \cos \phi \ket{g}, \\
\ket{-} &= \sin \theta \cos \phi \ket{0} + \cos \theta \cos \phi \ket{1} - \sin \phi \ket{g}, \\
\ket{d} &= \cos \theta \ket{0} - \sin \theta \ket{1},
\end{align*}
where the mixing angles $\theta$ and $\phi$ are given by
$\theta(t) = \tan^{-1}(\Omega_0(t)/\Omega_1(t))$ and
$2\phi(t) = \tan^{-1}\left(\Omega_{\text{rms}}(t)/\Delta\right)$ and $\Omega_{\text{rms}}(t)=\sqrt{|\Omega_0(t)|^2 + |\Omega_1(t)|^2}$.
The corresponding eigenfrequencies are given by,
$\varepsilon_{\pm}(t) = \frac{1}{2} \left( \Delta \pm \sqrt{\Delta^2 + \Omega_{\text{rms}}(t)^2 }\right)$, and $\varepsilon_{d}(t) = 0$.
The eigenstate with zero eigenenergy, the dark state $\ket{d}$
is the key to understand the transfer process.
When the system 
is initialised in state $\ket{0}$, the dark state $\ket{d}$
can be evolved adiabatically by changing the mixing angle 
$\theta(t)$ from 0 to $\pi/2$, as shown in 
Fig.~\ref{fig:fig1_stirap_concept}(c). 
It is important to note that when the single-photon detuning 
$\Delta$ and two-photon detuning $\delta$ is set to zero,
it leads to a energy gap of $\Omega_{\text{rms}(t)}/2$ between 
the instantaneous eigenenergies of the dark state and the two 
other states.
This gap facilitates adiabatic evolution through the dark 
state during the process, ultimately leading to population 
transfer from state $\ket{0}$ to state $\ket{1}$ without 
populating the intermediate state $\ket{g}$.
If the system is initialized in $\ket{1}$,  
the same pulse sequence results in an evolution through
an equal superposition of the $\ket{+}$ and $\ket{-}$ 
eigenstates.
Due to this, the final state is very sensitive to 
the phase acquired by these two eigenstates during 
the evolution, making it difficult to achieve successful 
robust state transfer when the system is initialized 
in the state $\ket{1}$, as shown in 
Fig.~\ref{fig:fig1_stirap_concept}(d). 
Thus, while the STIRAP protocol can be employed for 
directional state-transfer, it cannot work as a quantum 
gate between $\ket{0}$ and $\ket{1}$ due to its sensitivity 
to the initial state \cite{genov_robust_2023}. 

\subsection{Implementation of single-qubit gates using detuned STIRAP}

\begin{figure*}
\includegraphics[width=155mm]{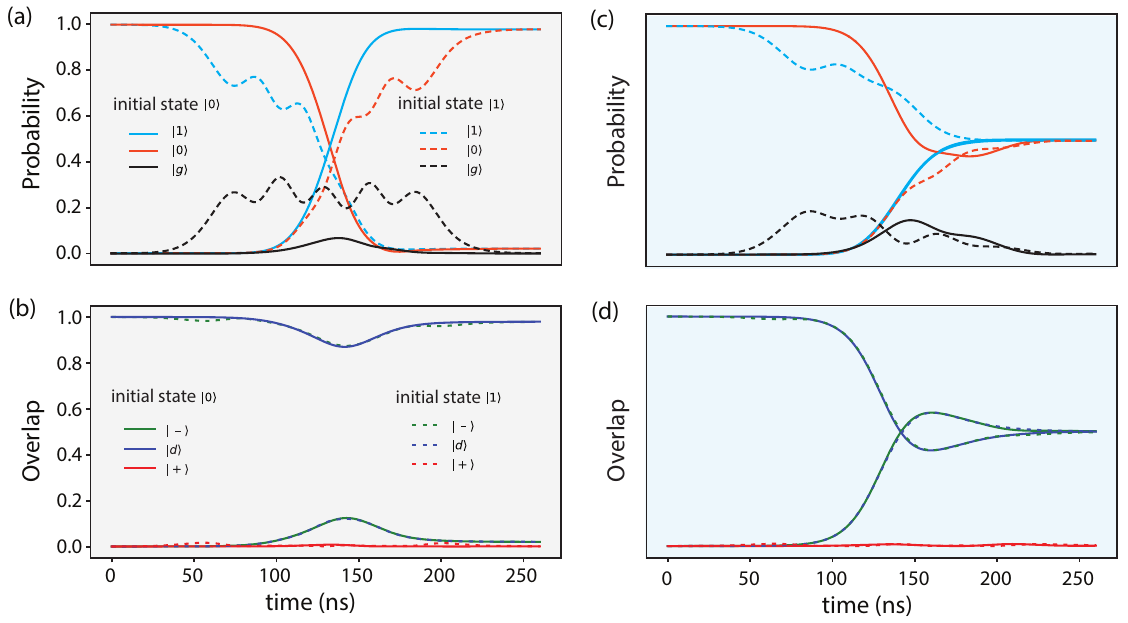}
\caption{\textbf{Time evolution for detuned STIRAP}: Evolution of states for the $\pi$ gate  
and $\pi/2$ gate with non-zero single photon 
detuning ($\delta = 0$ MHz, $\Delta = 15$ MHz). 
The figures are plotted for two different initial states: 
solid lines represent evolution from the $\ket{0}$ initial 
state and dashed lines represent an evolution from 
the $\ket{1}$ initial state.
(a) The population in different states during 
the evolution for the $\pi$ rotation. 
(b) The overlap of the evolving state with 
the instantaneous eigenbasis for the $\pi$ rotation. 
The evolution proceeds almost completely through 
a single eigenstate, $\ket{d}$ for the initial 
state $\ket{0}$ and $\ket{-}$ for the initial 
state $\ket{1}$.
(c) The population in different states during 
the evolution for the $\pi/2$ rotation. (d) The evolution does not 
proceed through a single eigenstate for the $\pi/2$ rotation; instead, there is mixing between 
the eigenstates $\ket{d}$ and $\ket{-}$, 
making the $\pi/2$ rotation only partially adiabatic 
with adiabaticity being maintained with 
respect to the $\ket{+}$ eigenstate.
\label{fig:eigen}}
\end{figure*}

The limitation of resonant STIRAP as a quantum gate 
arises from the initial degeneracy between the 
instantaneous eigenstates $\ket{+}$ and $\ket{-}$, 
resulting in an equal overlap with these states 
when initialized in $\ket{1}$.
To implement a STIRAP-inspired gate, this degeneracy 
can be partially removed by introducing moderate 
single-photon detunings. 
As a result, the state $\ket{+}$ can be adiabatically 
eliminated, and the final state evolves through 
$\ket{d}$, or $\ket{-}$ (or both). 
When starting in $\ket{0}$, the adiabatic evolution 
proceeds via the eigenstate $|d\rangle$, as per the 
usual STIRAP evolution.
However for $\ket{1}$, the evolution takes place
via $\ket{-}$ eigenstate, while keeping the overlap with 
$\ket{+}$ suppressed.
In this section, we give the theoretical framework 
of detuned STIRAP to better understand the $\pi$ 
and $\pi/2$ rotations between the state $\ket{0}$
and $\ket{1}$.
To begin, we transform $H(t)$ to the 
adiabatic basis using the rotation matrix,
\begin{align}
R(t)=\left(\begin{array}{ccc}
\sin \theta(t) \sin \phi(t) & \cos \phi(t) & \cos \theta(t) \sin \phi(t) \\
\cos \theta(t) & 0 & -\sin \theta(t) \\
\sin \theta(t) \cos \phi(t) & -\sin \phi(t) & \cos \theta(t) \cos \phi(t)
\end{array}\right).
\end{align}
In this frame, we obtain the effective Hamiltonian as,
\begin{align}
\begin{split}
&H_{\text{ad}}(t) \\
&\quad= R(t) H(t) R^{\dagger}(t)-i R(t) \partial_t R^{\dagger}(t) \\
&\quad=\begin{pmatrix}
\epsilon_+ (t) & \dot{\theta}(t)\sin\phi(t) & i\dot{\phi}(t) \\
-i\dot{\theta}(t)\sin\phi(t) & 0 & -i\dot{\theta}(t)\cos\phi(t) \\
-i\dot{\phi}(t) & i\dot{\theta}(t)\cos\phi(t) & \epsilon_- (t)
\end{pmatrix}.
\end{split}
\label{Eq:H_ad}
\end{align}
Following the  adiabatic theorem, the condition of adiabaticity,
$\braket{\psi_i|\dot{\psi}_j}\ll|\epsilon_i-\epsilon_j|$,
leads to the following conditions:

\begin{align}
\label{cond}
\begin{aligned}
    &\left.
    \begin{aligned}
        & |\dot{\phi}(t)| \ll |\epsilon_+(t)-\epsilon_-(t)| \\
        & |\dot{\theta}(t)\sin(\phi (t))| \ll |\epsilon_+ (t)|\\
        & |\dot{\theta}(t)\cos(\phi (t))| \ll |\epsilon_- (t)|
    \end{aligned}
    \right\}
\end{aligned}
\end{align}
where $\psi_{i}$ represents one of the instantaneous 
eigenstates $\{\ket{+},\ket{-},$ or $\ket{d}\}$.
To ensure the entire process remains adiabatic, all the 
conditions described in Eq.~\ref{cond} must be satisfied 
throughout the evolution.

\subsubsection{$\pi$ rotations}

To implement $\pi$ rotations with respect to $x$ or 
$y$-axes in the qubit subspace, we use the 
detuned STIRAP pulses by introducing a moderate 
single photon detuning. It leads to the adiabatic elimination 
of state $\ket{+}$.
Now, let’s consider the application of the detuned STIRAP 
pulses for two initial states $\ket{0}$ and $\ket{1}$. 
For the initial state $\ket{0}$, the instantaneous dark 
eigenstate at time $t=0$ aligns perfectly with it. 
During the evolution under the detuned STIRAP pulses,
the process remains nearly adiabatic, and the initial state 
evolves along $\ket{d}$ and ends up in $\ket{1}$ without 
populating the ground $\ket{g}$.
Similarly, for the initial state $\ket{1}$, the 
instantaneous eigenstate $\ket{-}$ aligns with it at 
time $t=0$. 
Again, as the process is nearly adiabatic, the initial 
state will evolve along $\ket{-}$ and end up at $\ket{0}$.
Fig.~\ref{fig:eigen}(a) shows the results from numerical
calculations based on QuTip \cite{johansson_qutip_2013},
showing the probabilities in different eigenstates as 
the state of the system evolves. 
Further, by using the time-dependent Hamiltonian in Eq.~\ref{eq:H0}, 
we found the instantaneous eigenstates ($\ket{d(t)}$, 
$\ket{-(t)}$ and $\ket{+(t)}$) at each point in time 
in the protocol. Afterwards, we evolve the initial states $\ket{0}$ and $\ket{1}$ with the protocol and found 
the overlap $|\braket{i(t) | \psi(t)}|^2$ during the 
evolution of the state, 
where $\ket{i(t)}\in\{\ket{d(t)},\ket{-(t)},\ket{+(t)}\}$, 
and $\psi(t)$ is the system's state during evolution.

Fig.~\ref{fig:eigen}(b) shows the results for two initial
states $\ket{0}$ and $\ket{1}$. The evolution proceeds 
through the dark-state $\ket{d}$ and $\ket{-}$. 
We further emphasize that due to the separation of 
the eigenenergies, the process remains nearly 
adiabatic. The small dip during the evolution is due to the finite length of the process. 
Thus, as far as the transfers $\ket{0}\leftrightarrow \ket{1}$  
are concerned, detuned STIRAP pulses perform a $\pi$ rotation in 
the qubit subspace about some axis in the $xy$-plane.

In case of $\pi$ rotations, it is also possible to 
analytically obtain the corresponding unitary. 
Using pulse parameters such 
that $\Omega_0=\Omega_1 \lesssim \Delta$,
the instantaneous eigenenergies always 
satisfy the adiabatic conditions given by Eq.~\ref{cond}.
It implies that the off-diagonal coupling terms in 
$H_\text{ad}$ are much smaller than the diagonal terms, 
thus leads to
\begin{align}
H_{\text{ad}}(t) \simeq \begin{pmatrix}
\epsilon_+ (t) & 0 & 0 \\
0 & 0 & 0 \\
0 & 0 & \epsilon_- (t)
\end{pmatrix}.
\end{align}

Given the diagonal form of $H_{\text{ad}}(t)$, we can write
$U(t,0) = R^\dag (t) e^{-i \int_{0}^{t} H_\text{ad}(t')\, dt'} R(0)$,
resulting in, 

\begin{align}
U(t,0) = \begin{pmatrix}
0 & 0 & e^{-i \eta_-} \\
0 & e^{-i \eta_+} & 0 \\
-1 & 0 & 0 \label{unitp}
\end{pmatrix}.
\end{align}
% The phase factors in the above unitary are given by $ \phi_+ = \int_{0}^{t} \epsilon_{+}(t') \, dt' $ and $ \phi_- = \int_{0}^{t} \epsilon_{-}(t') \, dt' $ .
The phase factors in the above unitary are given by  \( \eta_- = \int_{0}^{t} \epsilon_{-}(t') \, dt' \) and \( \eta_+ = \int_{0}^{t} \epsilon_{+}(t') \, dt' \).

The unitary in Eq.~(6) represents a $\pi$ rotation along an axis in the $xy$-plane. The specific axis of rotation can be controlled by adjusting the phases of the individual pulse envelopes. In particular, we modify the phase of both the detuned STIRAP pulse envelopes simultaneously: $\Omega_0(t) \rightarrow e^{i\beta} \Omega_0(t)$ and $\Omega_1(t) \rightarrow e^{i\beta} \Omega_1(t)$, where $\beta$ is the phase shift applied to both pulses.

By tuning this phase $\beta$, we can align the axis of the $\pi$ rotation to any desired direction in the $xy$-plane. This allows for the rotation to occur along the $x$-axis, $y$-axis, or any intermediate axis, depending on the chosen phase. Simultaneously adjusting the phases of the detuned STIRAP pulses in this way compensates for the accumulated dynamical phase, thus enabling precise control over the $\pi$ rotations about the $x$ or $y$-axes.

\subsubsection{$\pi/2$ rotations}

Similar to the $\pi$ rotations, $\pi/2$ rotations can also 
be implemented using the detuned STIRAP pulses. 
Using the pulse parameters $\Omega_0 = \Omega_1 <\Delta$, 
the state $\ket{+}$ can be adiabatically eliminated due to 
single-photon detuning.
However, these pulse parameters do not comply with the 
last adiabatic condition in Eq.~\ref{cond} and results 
in a partially adiabatic process.
To understand $\pi/2$ rotations, we consider the time evolution of states, $\ket{0}$ and $\ket{1}$ under the application of detuned STIRAP pulses.
These two states are aligned to either the dark state
$\ket{d}$ or $\ket{-}$ at time $t=0$. 
As the process is partially adiabatic, the initial 
state evolves involving both $\ket{d}$ and $\ket{-}$.
Regardless of the initial state, under a suitable choice
of pulse amplitude and equal superposition of $\ket{0}$ 
and $\ket{1}$ can be reached as shown in 
Fig.~\ref{fig:eigen}(c). 
It implies that the pulse sequence may perform a $\pi/2$ rotation along some axis in the $xy$-plane in the qubit subspace.
Similar to the $\pi$ rotations, the axis of the $\pi/2$ rotations can also be aligned along the $x$-axis or $y$-axis 
by adjusting the relative phase difference of the individual 
detuned STIRAP pulses.
As the process is not completely adiabatic, some 
of the off-diagonal terms in $H_\text{ad}$ 
are not negligible when comparing to diagonal terms
and Eq.~\ref{Eq:H_ad} becomes,
\begin{align}
H_{\text{ad}}(t) \simeq \begin{pmatrix}
\epsilon_+ (t) & 0 & 0 \\
0 & 0 & -i\dot{\theta}(t)\cos\phi(t) \\
0 & i\dot{\theta}(t)\cos\phi(t) & \epsilon_- (t)
\end{pmatrix}
\end{align}
The non-negligible off-diagonal terms complicate 
the analytical calculation of unitary as $H_{\text{ad}}(t)$ does not commute 
with itself at different times.
To get more details, we also compute the time evolution of instantaneous eigenstates. As expected, the evolution does not proceed through a 
single eigenstate, but there is a mixing between the 
eigenstates $\ket{d}$ and $\ket{-}$ as shown in Fig.~\ref{fig:eigen}(d).
Therefore, the $\pi/2$ rotation protocol is only partially 
adiabatic, where adiabaticity is maintained with respect 
to the $\ket{+}$ eigenstate.

\section{Experimental results}

\subsection{Device concept and basic characterization}

\begin{figure}
\centering
\includegraphics[width=85mm]{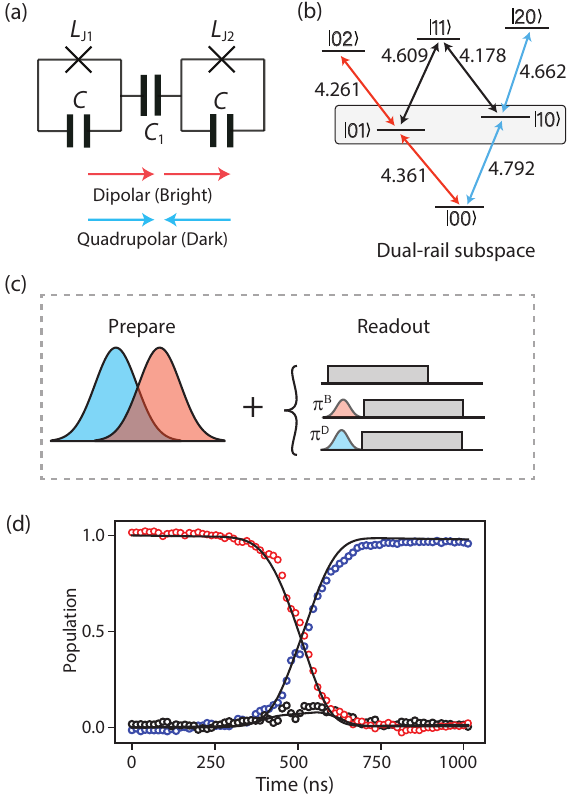}
\caption{\label{fig:device}(a):  
\textbf{Device concept and resonant-STIRAP:}
(a) Schematic of the device showing two nearly resonant 
capacitively coupled transmons. The strong coupling 
between the two transmons results in a dipolar and a quadrupolar modes.
(b) An experimentally determined energy-level diagram
showing different transitions in the 2-qubit manifold. 
The orange(cyan) arrows depict the dipolar (quadrupolar)
mode. The single excitation states in dipolar and quadrupolar
modes \textit{i.e.} $\ket{01}$ and $\ket{10}$ along with
the ground state $\ket{00}$($\ket{g}$) form a ``V-system" configuration 
for the coupled transmon system. The dual-rail subspace
is spanned by $\ket{01}$($\ket{0}$) and $\ket{10}$($\ket{1}$).
(c) The pulse protocol scheme for the demonstration 
of resonant-STIRAP in the time domain. 
The preparation pulse length is varied by truncating
it at different times before the readout. For a given 
preparation, three separate measurements are done to 
determine the populations in three levels. 
(d) Time evolution of the populations of different 
levels during resonant STIRAP, showing the transfer of 
population from $\ket{0}$ to $\ket{1}$.
}
\end{figure}

Our device consists of two strongly coupled 
nearly identical transmon qubits, which results 
into one dipolar mode and one quadrupolar
mode as shown schematically in Fig.~\ref{fig:device}(a).
The device is fabricated on a sapphire substrate using Tantalum-based
capacitor pads. Details of the device fabrication
and measurement apparatus are included in the 
supplemental materials. 
For the readout of these two-qubit modes, we use 
a 3D cavity and align the dipolar mode in the direction 
of the electric field of the fundamental cavity 
mode.
Due to this alignment, the dipolar mode couples
strongly to the cavity field, resulting in ``bright mode" 
(called qubit-$B$ from here on). However, the quadrupolar 
mode remains protected from the cavity field in 
the first order (called as ``Dark" or qubit-$D$ from 
hereon). 
A strong $ZZ$-coupling between the two modes results
in different dispersive shifts for the 
two modes and allows us to perform joint state 
readout \cite{bianchetti_control_2010}.
We begin with single-tone cavity spectroscopy and 
determine the dressed cavity frequency $\tilde{\omega}_{r}/2\pi=7.215$~GHz 
and a linewidth of $\kappa/2\pi\simeq1$~MHz. 
We perform two-tone spectroscopy to determine 
all the transitions in two coupled transmon 
manifold as shown in Fig.~\ref{fig:device}(b). 
From two-tone spectroscopy, we determine the 
anharmonicities of both modes to be $\alpha_{B}/2\pi\simeq100$~MHz 
and $\alpha_{D}/2\pi\simeq130$~MHz, respectively.
The time-domain characterization yields the coherences of 
$T_{1} = 64~\mu$s and $T_{2}^R =~106~\mu$s, for the B-mode,
and 
$T_{1} = 88~\mu$s and $T_{2}^R = 98~\mu$s for D mode.
A detailed characterization of the device is included in the 
supplemental material.

\subsection{Resonant STIRAP}

We begin with the demonstration of resonant STIRAP 
for the population transfer from $\ket{0}$ to $\ket{1}$ 
within the 3 lowest levels of the system, as shown in 
Fig.~\ref{fig:device}(b). 
First, the state $\ket{0}$ is prepared using $\pi$ pulse in the 
qubit-B manifold. For the resonant-STIRAP demonstration, we 
use two Gaussian pulses, each with a duration of 825 ns and 
a standard deviation of 133 ns. These pulses have equal 
amplitudes ($\Omega_0=\Omega_1$) and a time-offset
of 206~ns. Here, time offset denotes the time delay between the Gaussian pulses. To measure the time evolution of the population, 
these pulses can be truncated and immediately followed by
a readout pulse. 
To measure the populations in the higher levels $\ket{0}$, 
and $\ket{1}$, we transfer the population from these states 
to the ground state $\ket{g}$ using $\pi$ pulses on qubit-B 
and qubit-D, respectively \cite{bianchetti_control_2010}. 
The pulse sequence schematic is shown in Fig.~\ref{fig:device}(c).
The results from these measurements are shown
in Fig.~\ref{fig:device}(d). 
For the resonant-STIRAP composite pulses of duration 
of 1~$\mu$s, we observe that the transfer process is
not quite adiabatic, ending up populating $\ket{g}$
state partially during the transfer.
To understand our observations, 
we numerically model the system formed by three-levels
and based on the experimental parameters.
In these calculations, the decoherence errors are
included as they become non-negligible due to the 
longer pulse length.  
We find that nearly 98~\% population transfer and 
approximately 2~\% population decay to ground state 
due to relaxation during the evolution.

\subsection{Detuned STIRAP}
\begin{figure}
\includegraphics[width=75mm]{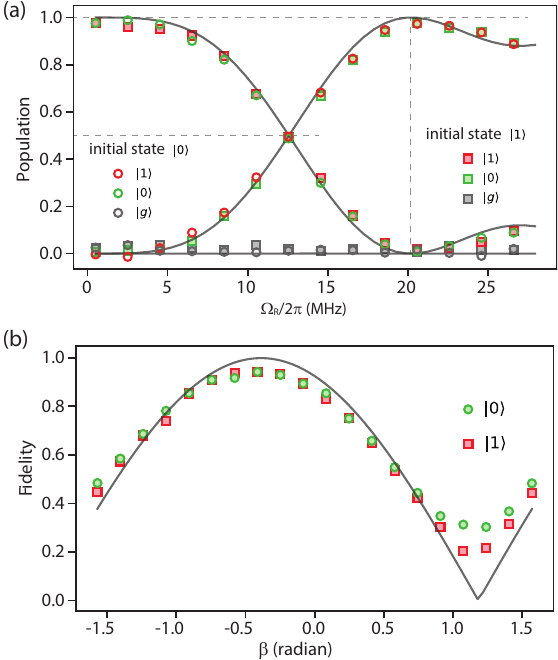}
\caption{\label{fig:ampsweep}
\textbf{Calibration of the $\pi$ and $\pi/2$ rotations:}
(a) Populations in the three states while varying the amplitude
$\Omega_R=\Omega_{0}=\Omega_1$ of the detuned STIRAP pulses. 
Results for both the initial states $\ket{0}$ and $\ket{1}$ are shown. 
For these measurements, a detuned STIRAP pulse sequence of 
260~ns duration is applied. 
(b) shows the calibration for the removal of the dynamical phase.
The overlap of prepared state with $(\ket{0}+\ket{1})/\sqrt{2}$
is plotted while varying the phase of detuned STIRAP 
pulses by the same amount simultaneously.
Solid-lines are numerically computed results using experimentally 
determined device and pulse parameters. Note that 
the solid-lines corresponding to each initial state lie
on top of each other and are not separately visible.}
\end{figure}

To demonstrate detuned STIRAP for $\pi$ and $\pi/2$ rotations,
we begin by the phase and amplitude calibrations. 
We set a moderate single-photon detuning of 15~MHz. 
A pulse configuration, schematically 
shown in Fig.~\ref{fig:fig1_stirap_concept}(b), having duration
of 206~ns, $\sigma=33~$ns, offset of 54~ns and equal 
Rabi amplitudes are used. 
We first prepare one of two initial states, $\ket{0}$ or $\ket{1}$.
It is followed by a detuned STIRAP pulse sequence and then
the sequences measuring the populations in all the states 
(as described earlier). Such a sequence is repeated
while varying the amplitudes of the detuned STIRAP
pulses.
Fig.~\ref{fig:ampsweep}(a) shows the experimental results 
showing that with an amplitude of approximately 20~MHz, 
one can successfully swap the populations 
from $\ket{0}\rightarrow\ket{1}$ and $\ket{1}\rightarrow\ket{0}$ 
using the same pulse amplitude and without populating 
the state $\ket{g}$.
Such a calibration, therefore can be used to implement 
$\pi$ rotations in the qubit-subspace.

We further note that the intersection of the two curves 
showing the transfer for the two initial states 
($\ket{0}$ and $\ket{1}$) takes place at a population of 0.5. 
It immediately suggests that half rotations
can be implemented by simply controlling the amplitude
of detuned STIRAP pulse.
To correct any dynamical phase acquired during the 
detuned STIRAP process and align the axes of rotation
to the lab-frame, we next vary the phase of the microwave
drives. The pulse amplitude is kept fixed 
to value corresponding to the common intersection point
of half the population.
It represents a superposition state with equal 
probability amplitudes \textit{i.e.}
$(\ket{0}+e^{i\gamma}\ket{1})/\sqrt{2}$,
where $\gamma$ is the dynamical phase that needs to be determined and corrected to get the desired quantum state.
We then perform quantum state-tomography of the 
prepared state (detailed in the next section) and compute
its overlap with $(\ket{0}+\ket{1})/\sqrt{2}$.
Fig.~\ref{fig:ampsweep}(b) shows the resultant 
fidelity as the phase of the microwave pulses is 
varied. The phase corresponding to the maximum fidelity represents a $\pi/2$ rotation about the $y-$axis. 
Thus, from these two calibrations, we can completely 
fix the amplitudes and phases required
to implement half- and full-rotations about the
$x-$, and $y-$axes on the Bloch-sphere spanned in the 
dual-rail subspace.
The solid lines in Fig.~\ref{fig:ampsweep} are 
numerically calculated using the experimentally
determined device and pulse parameters.

\subsection{Quantum state tomography (QST) of entire ``V-shape" system}

\begin{figure*}
\includegraphics[width=125mm]{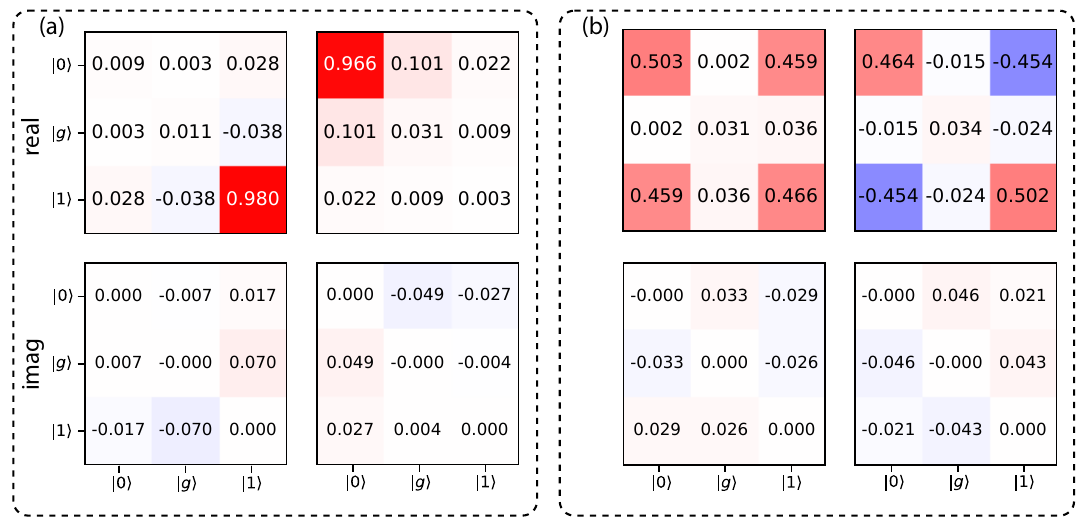}
\caption{\label{fig:QST} \textbf{Quantum state tomography for different initial 
states $\ket{0}$ (left) and $\ket{1}$ (right) for both $\pi$ and $\pi/2$ gates:} a) Real and imaginary plot of density matrix for $\pi$ gate. b) Real and imaginary plot of density matrix for $\pi/2$ gate. The fidelities of the four target states are
0.990, 0.983, 0.971, and 0.967, each with an error margin of 0.7\% in measurement.  }
\end{figure*}

To further benchmark and check the
completeness of the gates constructed using $\pi$ 
and $\pi/2$ rotations, we experimentally measure 
the complete density matrix corresponding to the $3-$level system
using three-level quantum state tomography \cite{bianchetti_control_2010}.
To perform a comprehensive QST, a complete set of 
nine independent measurements are carried out after 
preparation of a given state and calculating the 
density matrix based on the measurement outcomes. 
Before the measurements, we rotate the state by 
applying 9 different rotations consisting of different 
$\pi$ 
and $\pi/2$ rotations \textit{i.e.},
$\hat{R}_k = \{I,\left(\frac{\pi}{2}\right)^{g0}_x$, $\left(\frac{\pi}{2}\right)^{g0}_y$, $\left(\pi\right)^{g0}_x$, $\left(\frac{\pi}{2}\right)^{g1}_x$, $\left(\frac{\pi}{2}\right)^{g1}_y$, $\left(\pi\right)^{g0}_x\left(\frac{\pi}{2}\right)^{g1}_x$, $\left(\pi\right)^{g0}_x\left(\frac{\pi}{2}\right)^{g1}_y$, $\left(\pi\right)^{g0}_x\left(\pi\right)^{g1}_x \}$, 
where $I$ denotes the identity and 
$\left(\theta\right)^{ij}_a$ denotes $\theta$-rotation 
about $a$-axis between $i,j$ transitions.
The measurement operator is given by
$\hat{M}_{i} = \alpha_g(t)|g\rangle\langle g|+ \alpha_0(t)|0\rangle\langle0|+\alpha_1(t)1\rangle\langle1|$,
where $\alpha_i(t)$ denotes the averaged 
transmitted field amplitudes for the state $i$. 
For each of these unitary rotations, we measure 
the coefficients $\langle I_k \rangle=\text{Tr}(\rho R_k \hat{M}_i \hat{R}_k^\dagger)$ 
by integrating the phase quadrature of the transmitted signal 
over the measurement time $t$. We then solve the 
linear equations $\text{Tr}(\rho_\text{est} R_k \hat{M}_i \hat{R}_k^\dagger) = \langle I_k\rangle _\text{meas}$ to reconstruct the density matrix $\rho_\text{est}$. 
The reconstructed density matrices $\rho_\text{est}$ 
are post-processed using a maximum likelihood 
estimation procedure \cite{filipp_two-qubit_2009,james_measurement_2001}.
Fig.~\ref{fig:QST} summarizes the experimentally 
obtained density matrices for four different states.
Experimentally,  we first initialize the system in 
state $\ket{0}$ or $\ket{1}$ by exciting it from the
ground state. 
It is followed by a calibrated detuned STIRAP pulse 
corresponding to $\pi$ or $\pi/2$ rotations, and 
subsequently by the state tomography sequences.
When starting with $\ket{0}$ or $\ket{1}$, the system ends 
up in state $\ket{1}$ or $\ket{0}$ when a detuned STIRAP 
based $\pi$ rotation about the $y$-axis is applied. 
Similarly, equal superposition states are prepared 
with the application of $\pi/2$ rotation about the $y$-axis
while initializing the system in $\ket{0}$ or $\ket{1}$.

\subsection{Robustness of the detuned STIRAP based rotation}

\begin{figure}
\includegraphics[width=65mm]{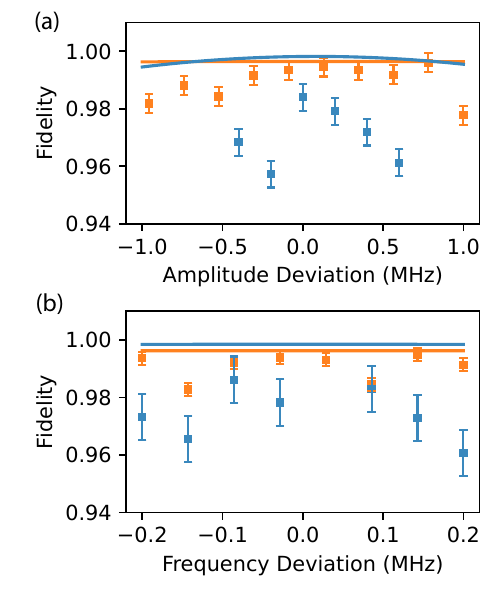}
\caption{\label{fig:robustness}
\textbf{Robustness of $\pi$ and $\pi/2$ rotations with respect to pulse amplitude and detuning:}
The solid lines represents the numerical simulations and 
scatter points show the experimental data. (a) Robustness 
of gates with respect to amplitude error for $\pi$ (orange)
and $\pi/2$ (blue). (b) Robustness of gates with respect to 
detuning error for $\pi$ (orange) and $\pi/2$ (blue). }
\end{figure}

To demonstrate the robustness of rotations with respect to 
deviations in the pulse amplitude and frequency,
we target to prepare states $\ket{1}$ and $(\ket{0}+\ket{1})/\sqrt{2}$
by the application of $\pi$ and $\pi/2$ rotations about $y$-axis
after initializing the system to state $\ket{0}$.
We carry out the three-level quantum state tomography 
as described earlier and compute the fidelity of the 
prepared states while varying the pulse amplitudes and frequency 
and compare the results with numerical simulations based
on experimentally determined parameters.
Fig.~\ref{fig:robustness} summarizes the experimental 
and numerical results for both $\pi$ and $\pi/2$ rotations. 
The experimental data indicate that both rotations
are somewhat resilient to deviations in amplitude and 
frequency. Whereas the numerically simulated plots remain 
relatively flat with respect to small deviations 
in the pulse parameters. 
There are several potential sources for the discrepancy
between the experimental and theoretical results.
First, we use a 15~MHz single-photon detuning in the experiments, 
which introduces non-adiabatic effects and can degrade 
the target state fidelity.
Second, the errors corresponding to the state preparation 
and the errors in the calibration of the gates used 
in tomography also contribute.
For the experimental results shown in Fig.~\ref{fig:robustness},
we have not removed any state preparation, measurement,
and decoherence errors.
On the technical side, a single microwave vector signal 
generator has been used to generate the control pulses 
for both the STIRAP pulses using a single sideband 
modulation technique. Due to pulse distortion and 
different pulse offsets, the calibration of the different 
rotations could have small error, which ultimately 
lowers the target state fidelities.
Pulse optimization 
\cite{zheng_optimal_2022,vasilev_optimum_2009,brown_reinforcement_2021,zhang_high-fidelity_2024} 
and shortcut to adiabaticity techniques 
\cite{chen_shortcut_2010,mortensen_fast_2018} 
can be employed to achieve higher state 
fidelities.
Additional numerical analysis (detailed in the 
supplementary materials) reveals that employing 
a single-photon detuning greater than 15~MHz 
could result in a preparation state fidelity 
in excess of 0.999. 
Within this parameter range, the gates could perform 
approximately 5\% better than typical resonant 
dynamical gates.

To summarize, we demonstrate detuned STIRAP inspired 
single qubit rotations in a dual-rail qubit formed by 
two fixed frequency transmons. 
We show that these gates are resilient to amplitude 
and detuning errors simultaneously when compared to 
the resonant pulse based rotations.
While we use state-fidelity as the performance 
metric to check the robustness of these rotations 
with pulse amplitudes and frequency, it can be better 
understood by implementing quantum process tomography 
and randomized benchmarking protocols in the future. 

\subsection{Acknowledgment}
The authors acknowledge the support 
under the CoE-QT program by MEITY and QuEST program by DST, Govt. of India (GoI). 
The authors acknowledge device fabrication facilities at CeNSE, 
IISc Bangalore, and central facilities at the Department of 
Physics funded by DST (GoI).
U.S. acknowledge the support received through the 
PMRF (GoI). 
\section{Appendix A: Device}

\subsection{Summary of the key device parameters}

\begin{table}[h]
\centering
\renewcommand{\arraystretch}{1} % Adjust row height
\caption{Device parameters}
\label{tab:parameters}
\begin{tabular}{|l|l|l|l|}
  \hline
  \textbf{Parameter} & \textbf{Symbol} & \textbf{Value} & \textbf{Units} \\
  \hline
  Dressed cavity frequency & $\tilde{\omega}_{r}/2\pi$ & 7.205 & GHz \\
  \hline
  Cavity line width & $\kappa/2\pi$ & 1 & MHz \\
  \hline
  Bright mode qubit frequency & $\tilde{\omega}_{B}/2\pi$ & 4.361 & GHz \\
  \hline
  Dark mode qubit frequency & $\tilde{\omega}_{D}/2\pi$ & 4.792 & GHz \\
  \hline
  Relaxation time of bright mode & $T_{1}(\text{B})$ & 64 & $\mu$s \\
  \hline
  Relaxation time of dark mode & $T_{1}(\text{D})$ & 88 & $\mu$s \\
  \hline
  Ramsey fringe time of bright mode & $T_{2}^R(\text{B})$ & 106 & $\mu$s \\
  \hline
  Ramsey fringe time of dark mode & $T_{2}^R(\text{D})$ & 98 & $\mu$s \\
  \hline
  Anharmonicity of bright mode & $\alpha_b/2\pi$ & 100 & MHz \\
  \hline
  Anharmonicity of dark mode & $\alpha_d/2\pi$ & 130 & MHz \\
  \hline
  Longitudinal coupling & $g_{zz}/2\pi$ & 180 & MHz \\
  \hline
  Dispersive shift of bright mode & $\chi_b/2\pi$ & 1.2 & MHz \\
  \hline
  Dispersive shift of dark mode & $\chi_d/2\pi$ & 1.55 & MHz \\
  \hline
  Coupling of bright mode to cavity & $g_{b}/2\pi$ & 150 & MHz \\
  \hline
  Junction asymmetry parameter & $d_{J}$ & 0.01 & \\
  \hline
\end{tabular}
\end{table}

\subsection{Device fabrication}\label{fab}

\begin{figure*}
\includegraphics[width=160mm]{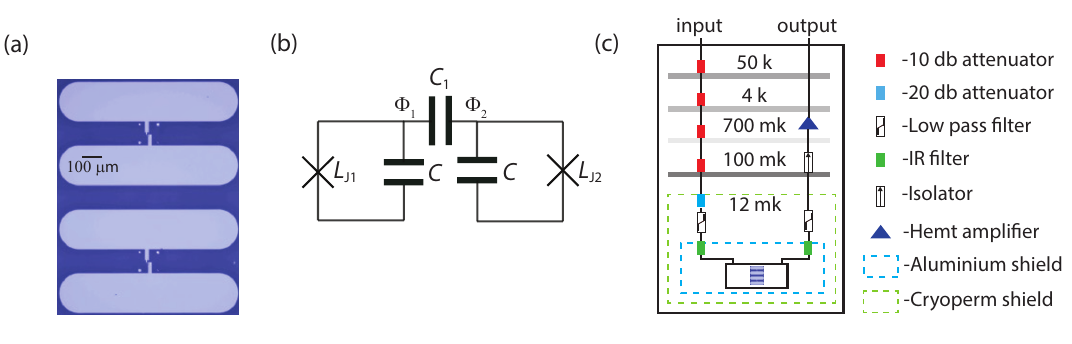}
\caption{\label{fig:supp_sch}\textbf{Device and measurement setup:} 
(a) An optical microscope image of a coupled transmon device. 
(b) An effective lumped element model of the two qubits showing node fluxes. 
(c) A schematic of the measurement setup showing various microwave 
components on the input and output lines. The low pass tubular 
filter (K\&L 6L250-8000/T20000) and an isolator (LNF-ISC4\_8A) 
from Low Noise Factory are used.
Two stages of shielding protect the sample.
The first stage is an aluminum shield coated with a mixture of carbon lampblack and silicon carbide grit to protect the 
device against IR radiation. The second stage is a cryo-perm shield 
for protection against the magnetic field.}
\end{figure*}

The 3D transmon qubits are fabricated on $c$-plane 
double-side polished 330~$\mu$m thick sapphire substrates.
The transmon capacitor pads are made up of 
tantalum (Ta), and the Josephson junctions 
are fabricated using Al-AlO$_\text{x}$-Al.
We grow 45 nm Ta films using the DC-magnetron sputtering process.
We use an argon flow of 40~sccm, DC power of 250~W. 
The distance between the target and substrate holder 
is set at 180 mm, while the substrate holder is maintained 
at a temperature of 600\degree~C.
Additionally, the substrate rotation speed is set to 10 rpm. 
The sputtering pressure is maintained at 3~mTorr.
These specified conditions collectively contribute to 
the controlled and optimized sputtering deposition process.
For the fabrication of Josephson junctions, we employ 
a double-layer resist stack to craft Manhattan-style junctions.
Initially, we spin coat EL-11 and bake it for 5 minutes. 
Subsequently, we spin coat a second layer of ARP-6200 resist, 
followed by a 5-minute baking process. 
Finally, we top it off by coating a layer of a water-soluble 
discharging layer onto the resist stack to mitigate 
any potential charging issues during the lithography process.
After exposure, we remove the anti-charging layer 
by dipping the chip in DI water for 60 seconds. 
Next, we remove the exposed ARP-6200 resist by using 
CSAR as a developer and IPA as a stopper. 
Finally, we remove the exposed EL-11 by using 
1:3 ratio of MIBK:IPA as a developer and IPA as a 
stopper. 
After development, the oxygen plasma is carried 
out for 40 seconds to remove the residual resist. 
The chip is then loaded into the electron beam evaporator.
Before aluminum deposition, a crucial step is to 
remove the native oxide of tantalum through Ar-ion milling. 
The ion milling process involves specific parameters, 
including an Argon flow of 5 sccm, a beam voltage of 
400~V, an accelerating voltage of 90 V, a beam current 
of 15~mA, a neutralizer emission current of 15~mA, 
and a duration of 40~seconds.

Following the ion milling, the next step involves 
depositing 15 and 30~nm of aluminum from two different 
directions with an intermediate oxidation step 
(2 Torr for 20 minutes). After the completion of 
deposition, the chip is kept in N-Methylpyrrolidone (NMP) 
solution for 3 hours at a temperature of 80°C before 
performing the liftoff.
An optical microscope image of the fabricated device
is shown in Fig.~\ref{fig:supp_sch}(a).

\subsection{Measurement setup}
The samples are cooled down using a BlueFors BF-LD250 
dilution refrigerator with a base temperature of 12~mK. 
The samples are attached to a flange connected to the 
lowest temperature stage, which is directly cooled 
down by the dilution process in the mixing chamber.
Various components, such as attenuators, filters, amplifiers, 
and isolators at various stages are shown in Fig.~\ref{fig:supp_sch}(c).
The control pulse and the measurement pulse are
combined at room temperature, and the combined signal 
is sent to the input port of the cavity by using a 
single coaxial line.
The output measurement pulse from the cavity 
is first sent through HEMT, followed by another 
amplifier at room temperature.

\section{Appendix B: Hamiltonian of two strongly coupled transmon}
\label{sec:modes}

We start from the classical Lagrangian \(L\) of the 
circuit as shown in Fig.~\ref{fig:supp_sch}(b), showing 
the generalized flux variables at the left and right 
nodes of the circuit denoted by \(\Phi_1\) and 
\(\Phi_2\). The kinetic energy \(K\), stored in 
the capacitors can be written as,
\begin{align}
K = \frac{C}{2} \dot{\Phi}_1^2 + \frac{C}{2} \dot{\Phi}_2^2 + \frac{C_c}{2} \left(\dot{\Phi}_1 - \dot{\Phi}_2\right)^2
\end{align}
Similarly, we can write the potential energy $U$, 
stored in the junctions as,
\begin{align}
\begin{split}
U=-&E_J\left(1-d_J\right) \cos \left(\frac{\Phi_1}{\phi_0}\right)\\ -&E_J\left(1+d_J\right) \cos \left(\frac{\Phi_2}{\phi_0}\right),
\end{split}
\end{align}
where $d_J=\left(E_{J 2}-E_{J 1}\right)/\left(E_{J 2}+E_{J 1}\right)$ 
is an asymmetry parameter in the junctions
in the Junction and $E_J=\left(E_{J 2}+E_{J 1}\right) /2$ 
is the average Josephson energy, \(\Phi\) is the node flux 
and $\phi_0=\Phi_0 /2\pi=\hbar/2e$ is the reduced 
magnetic flux quantum. It is convenient to introduce 
``bright" and ``dark" variables \(\Phi_b\) 
and \(\Phi_d\) as the flux average 
$\Phi_b=\left(\Phi_1-\Phi_2\right) / 2$ and 
$\Phi_d=\left(\Phi_1+\Phi_2\right) / 2$. 
Now, the Lagrangian $L= K-U$ can be defined as,
\begin{align}
\begin{split}
L=C& \ \dot{\Phi}_d^2+\left(C+2 C_1\right) \dot{\Phi}_b^2 \\ +&2 E_J\left[\cos \left(\frac{\Phi_b}{\phi_0}\right) \cos \left(\frac{\Phi_d}{\phi_0}\right)\right] \\+&2 E_Jd_J\left[\sin \left(\frac{\Phi_b}{\phi_0}\right) \sin \left(\frac{\Phi_d}{\phi_0}\right)\right].
\end{split}
\end{align}
We now calculate the conjugate charges $Q_b$ 
and $Q_d$, corresponding to the phases 
$\Phi_b$ and $\Phi_d$ as,
\begin{align}
Q_b=\frac{\partial\mathrm{L}}{\partial\dot{\Phi}_b}=(2 C +4C_1) \dot{\Phi}_b
\end{align}
\begin{align}
    Q_d=\frac{\partial \mathrm{L}}{\partial \dot{\Phi}_d}=2 C \dot{\Phi}_d
\end{align}
Now, we can obtain the Hamiltonian of the 
circuit \[H(Q_b, Q_d, \Phi_b, \Phi_d) = Q_b \dot{\Phi}_b + Q_d \dot{\Phi}_d - L\] as,
\begin{align}
\begin{split}
H=&\frac{Q_b^2}{2 C_b}+\frac{Q_d^2}{2 C_d}-2 E_J \cos \left(\frac{\Phi_b}{\phi_0}\right) \cos \left(\frac{\Phi_d}{\phi_0}\right) \\ &-2 E_J d_J \sin \left(\frac{\Phi_b}{\phi_0}\right) \sin \left(\frac{\Phi_d}{\phi_0}\right),
\end{split}
\end{align}
where the effective capacitances are $C_b = 2C + 4C_1$ and $C_d = 2C$,
respectively. We can now quantize this Hamiltonian by promoting the
flux and charge variables to operators. 
In addition, we define dimensionless phase operators 
$\hat{\varphi}_j=\hat{\Phi}_j / \phi_0$ and 
charge number operators $\hat{n}_j = \hat{Q}_j/2e$, 
and use them to express the quantum Hamiltonian 
of the circuit as,
\begin{align}
\begin{split}
    H =&\ 4\hat{n}_b^2E_{C_b} + 4\hat{n}_d^2E_{C_d} - 2E_J \cos\hat{\varphi}_b \cos\hat{\varphi}_d \\ &- 2E_J d_J \sin\hat{\varphi}_b \sin\hat{\varphi}_d.
\end{split}
\end{align}
We define $E_{C_b} = e^2/(2C_b)$ and $E_{C_d} = e^2/(2C_d)$ 
as the charging energy corresponding to the dark 
and bright mode, respectively.
In the low anharmonicity limit, $\hat{\phi}_a, \hat{\phi}_b \ll 1$, 
we can expand the cosines and retain the terms up to 4$^{\text{th}}$ 
order in the phases. 
Also, if the junctions are nearly 
identical $d_J \ll 1$, we can simplify the Hamiltonian by 
taking $d_J = 0$ as it will only add small unwanted 
coupling terms between the qubits \cite{richer_inductively_2017}. 
By taking into account the above approximations 
we can finally write the Hamiltonian as,
\begin{align}
\begin{split}
    \hat{H} =&\ 4E_{C_b} \hat{n}_b^2 + \frac{E_{J_b}}{2} \hat{\varphi}_b^2- \frac{E_J}{12} \hat{\varphi}_b^4 + 4E_{C_d} \hat{n}_d^2\\+& \frac{E_{J_d}}{2} \hat{\varphi}_d^2  - \frac{E_J}{12} \hat{\varphi}_d^4 - \frac{E_J}{2} \hat{\varphi}_b^2 \hat{\varphi}_d^2 + \mathcal{O}(6),\\
\end{split}
\end{align}
where $E_{J_b} =E_{J_d}=2E_J$. 
By defining the annihilation and creation 
operators corresponding to both the modes as follows,
\begin{align}
% \begin{split}
    \hat{\varphi}_b &= \left(\frac{8E_{C_b}}{E_{J_b}}\right)^{1/4}\frac{\hat{b} + \hat{b}^\dagger}{\sqrt{2}} \\
    \hat{n}_b &= -i \left(\frac{E_{J_b}}{8E_{C_b}}\right)^{1/4} \frac{\hat{b} - \hat{b}^\dagger}{\sqrt{2}} \\
    \hat{\varphi}_d &= \left(\frac{8E_{C_d}}{E_{J_d}}\right)^{1/4}\frac{\hat{d} + \hat{d}^\dagger}{\sqrt{2}} \\
    \hat{n}_d &= -i \left(\frac{E_{J_d}}{8E_{C_d}}\right)^{1/4} \frac{\hat{d} - \hat{d}^\dagger}{\sqrt{2}},
% \end{split}
\end{align}
we can reduce the Hamiltonian as,
\begin{align}
\begin{split}
    H =&  \
 \widetilde{\omega}_b b^\dagger b - \frac{\alpha_b}{2} b^\dagger b^\dagger bb + \widetilde{\omega}_d d^\dagger d \\&- \frac{\alpha_d}{2} d^\dagger d^\dagger dd  - 2g_{zz} b^\dagger b d^\dagger d,
\end{split}
\end{align}
% \end{minipage}
%
%
where $\alpha_b = E_{C_b}$, $\alpha_d = E_{C_d}$, 
$g_{zz} = \sqrt{E_{C_b}E_{C_d}}$, 
$\widetilde{\omega}_b = \sqrt{8E_{J_b}E_{C_b}} - \alpha_b - g_{zz}$,
and
$\widetilde{\omega}_d = \sqrt{8E_{J_d}E_{C_d}} - \alpha_d - g_{zz}$.
These expressions represent two transmon qubits 
which are strongly coupled to each other 
by a longitudinal coupling $g_{zz}$.
\subsection{Coupling of qubits with cavity}\label{sec:dimon description}
For the joint-state readout of both the ``qubits",
we couple them to a 3D microwave copper cavity. 
The coupled Hamiltonian of such a three mode system 
can be expressed as,
\begin{align}
\begin{split}
    H =&\ \omega_{r} a^\dagger a + \widetilde{\omega}_b b^\dagger b - \frac{\alpha_b}{2} b^\dagger b^\dagger b b + \widetilde{\omega}_d d^\dagger d \\&- \frac{\alpha_d}{2} d^\dagger d^\dagger d d - 2g_{zz} b^\dagger b d^\dagger d \\&+ g_b(a^\dagger b +a b^\dagger)+ g_d(a^\dagger d + a d^\dagger),
\end{split}
\end{align}
where $\omega_r$ represents the bare frequency of 
the readout cavity and $g_{b}(g_{d}$) represents 
its coupling to the bright (dark) mode.
The bright mode is attributed to the dipolar mode due 
to its alignment with the fundamental modes, the cavity
and it couples strongly to the cavity.
The dark mode is attributed to the quadrupole 
mode of oscillation, and it does not  
couple to the cavity as strongly.
Therefore, for the simplification, we set the dipolar 
coupling of the dark mode to zero \textit{i.e.} 
$g_{d} = 0$. 
After finding the eigenenergies using the non-degenerate 
perturbation theory \cite{zhang_suppression_2017}, the 
dispersive shifts corresponding to both modes 
can be calculated as,
\begin{align}
\chi_b = \frac{2g_{b}^2 \alpha_{b}}{\Delta_{b}(\alpha_{b} - \Delta_{b})}
\end{align}
\begin{align}
\chi_d =\frac{2g_{b}^2 g_{zz}}{\Delta_{b}(2g_{zz} - \Delta_{b})}
\end{align}

\begin{figure}
\includegraphics[width=70mm]{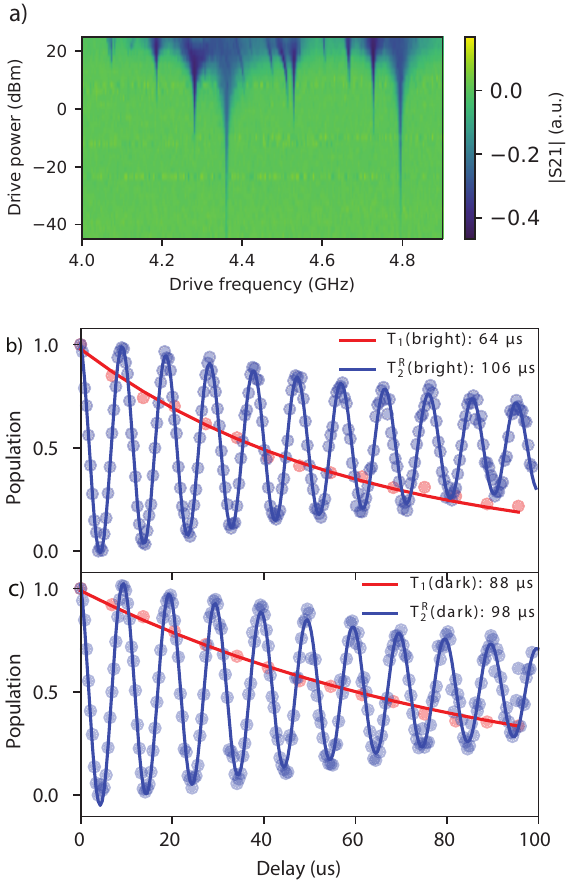}
\caption{\label{fig:supp_two_tone}\textbf{Qubits spectroscopy and coherence:} 
(a) Two tone spectroscopy showing the coupled qubit spectrum as 
the drive tone frequency and its strength are varied. Higher transitions
in the coupled qubit manifold (including multi-photon transitions
can also be seen. (b, c) Energy relaxation and Ramsey-fringe measurements 
for the bright and dark modes. }
\end{figure}

\begin{figure*}
\includegraphics[width=170mm]{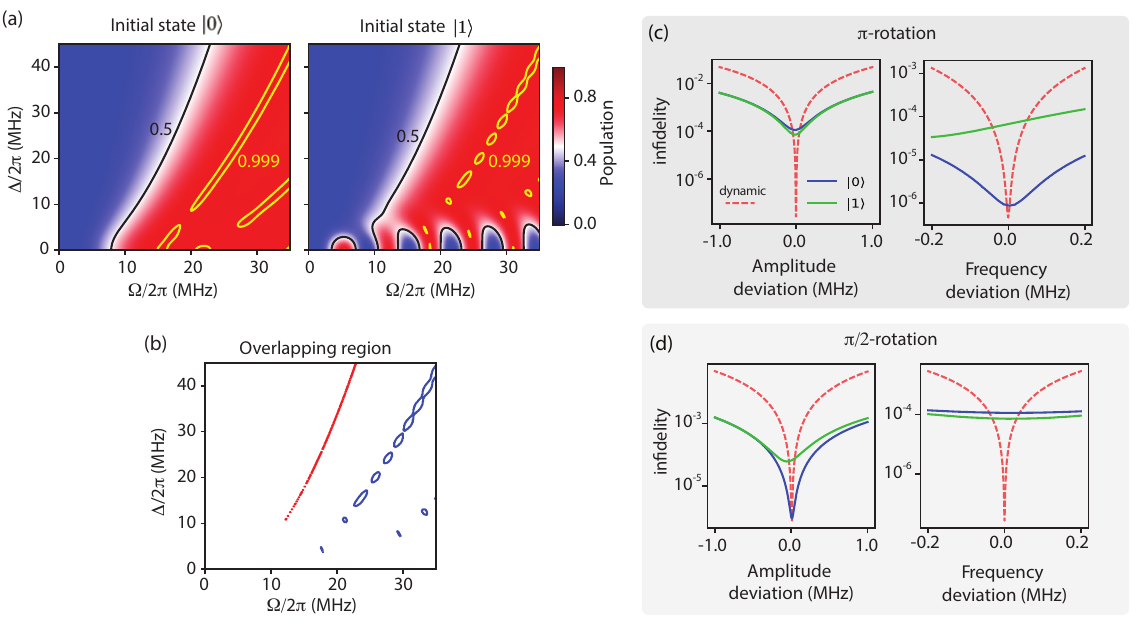}
\caption{\label{fig:supp_gate}
(a) Numerically calculated plot of the population in 
$\ket{1}$ (left panel) after the end of the detuned STIRAP 
pulse with an initial state of $\ket{0}$. 
The right panel shows the results from similar calculations 
and plot the population in $\ket{0}$ with an initial 
state of $\ket{1}$.
(b) Plot shows the common regions in the parameter
space for which detuned STIRAP protocol prepare
an equal superposition (red) or the complete transfer 
(blue) within an error tolerance of $10^{-3}$.
(c) Panel shows the numerically calculated state infidelity
for the $\pi$ rotation with respect to the 
deviation in the pulse amplitude and frequency around the 
optimum point. 
The blue (green) line represents the result when the system is 
prepared in the initial state $\ket{0} (\ket{1})$. The state 
infidelity is calculated using the target state $\ket{1} (\ket{0})$.
(d) Panel shows the state infidelity for the $\pi/2$ rotation with respect to deviation in amplitude/frequency 
around the optimum point. 
For an initial state of $\ket{0}$, the infidelity is computed 
using the target state of $(\ket{0}+\ket{1})/\sqrt{2}$ and is shown with blue line.
Similarly, the green line represents the result, 
when the system is initially prepared in state $\ket{1}$ 
and final state is measured in state $(\ket{0}-\ket{1})/\sqrt{2}$. 
For comparison, similar results are obtained by using 
dynamic rotations are also included (red-dashed line).   
}
\end{figure*}

It is interesting to point out that the dark (quadrupole) 
mode mostly remains decoupled from the cavity, therefore
does not directly contribute to the dispersive shift.
However, due to the strong cross-Kerr interaction between 
the qubit modes, the dispersive shift of the dark mode 
depends on the cross-kerr coupling.
In the dispersive limit, the Hamiltonian can be 
simplified to
\begin{align}
\begin{split}
    H = &\ \tilde{\omega}_{b}' {b}^\dagger b
     + \tilde{\omega}_{d} {d}^\dagger d + \tilde{\omega}_{r} a^\dagger a - \frac{\alpha_b}{2} b^\dagger b^\dagger b b \\&- \frac{\alpha_d}{2} d^\dagger d^\dagger d d+ \chi_{b} a^\dagger a b^\dagger b \\&+ \chi_{d} a^\dagger a d^\dagger d   - 2g_{zz} b^\dagger b d^\dagger d,
\end{split}
\end{align}
where $\tilde{\omega}_{b}' = \tilde{\omega}_{b}+\frac{g_{b}^2}{\Delta_{b}}$ and $\tilde{\omega}_{r} = \omega_{r}-\frac{g_{b}^2}{\Delta_{b}}$

\section{Appendix C: Device characterization}

To determine the transition frequencies for both 
qubit modes, we begin with the two-tone spectroscopy
method. The cavity is continuously probed at the 
dressed frequency of 7.215~GHz while varying
the frequency of an excitation tone applied near
the expected qubit transitions.
Fig.~\ref{fig:supp_two_tone}(a) show the results 
from the two-tone spectroscopy measurements. 
It is evident that at very low drive power, only 
two modes persist, revealing qubit mode frequencies 
of 4.361~GHz and 4.792~GHz. The drive power in 
Fig.~\ref{fig:supp_two_tone}(a) refers to the 
signal generator's output.
After the initial spectroscopic characterization, 
we carry out time domain coherence measurements. 
Fig.~\ref{fig:supp_two_tone} (b, c) shows the coherence 
measurements for both the qubit modes. We emphasize 
here that based on the simulations, the energy relaxation
rate for the bright mode is Purcell-factor limited. 
The dark mode remains uncoupled from the readout 
cavity, so it shows a somewhat larger energy 
relaxation time. In Table~\ref{tab:parameters}, we 
summarize the various parameters of the device.

\section{Appendix D: Numerical results of detuned STIRAP based rotations}
To study detuned STIRAP-based rotations to 
implement it as a quantum gate, we introduce 
single-photon detuning to lift the degeneracies 
in instantaneous eigenvalues. 
Using QuTiP, we solve the time-dependent 
Lindblad master equation to simulate the evolution 
of the system under the time-dependent Hamiltonian 
described by Eq.~1 of the main text. 
Our investigation focuses on understanding 
how single-photon detuning $\Delta$ and drive 
amplitude $\Omega$ affects the final state. 
We analyze this evolution across different 
parameter settings and the two initial states, 
namely $\ket{0}$ and $\ket{1}$.
We use Gaussian pulses of the total duration of 206~ns, 
standard deviation of $\sigma$ of 40~ns, an offset 
of 54~ns, and of equal amplitudes.

The left panel of Fig.~\ref{fig:supp_gate}(a) shows
the population in $\ket{1}$ for an initial state $\ket{0}$ 
when subjected to a 260~ns (206~ns + 54~ns) while varying the
single-photon detuning and amplitude.
Similarly, the right panel shows the population 
in $\ket{0}$ for an initial state $\ket{1}$. 
In these panels, the solid black lines indicate 
the contour corresponding to the population of 0.5. 
The yellow lines denote regions where the population 
at the end of the operation reaches 0.999. 

It is important to note that there are common regions
in both the panels \textit{i.e.} common values of the
single-photon detuning and amplitude, where the 
equal superposition or the complete transfer can
be achieved. Thus, these values of detuning and 
drive amplitude makes the rotations independent
of the initial condition. 
Fig.~\ref{fig:supp_gate}(b) illustrate the common 
regions in both plots with a tolerance 
of $10^{-3}$ for the equal superpositions and complete
transfer.
This figure delineate specific regions where detuned STIRAP can be utilized for $\pi/2$ and $\pi$ rotations.
Due to the finite-length of the detuned STIRAP pulses,
their rotations are not perfect. To better understand 
these imperfections and their sensitivity 
to pulse parameters, we carry out QuTiP-based 
numerical calculations.
Fig.~\ref{fig:supp_gate}(c, d) summarizes the results from
these calculations. The top panel shows the infidelity of 
the prepared state after the application of $\pi$ rotation
while varying the amplitude and frequency of the
detuned STIRAP pulses. 
Similar results for the $\pi/2$ rotations are shown
in the bottom panel.
For a comparison, we also plot the results for the 
standard dynamical rotations implemented by 
a resonant pulse driving the qubit subspace denoted by red dashed line.
It is evident from the figure that while dynamical 
rotations can offer much lower errors, these are
not enough robust with respect to frequency and 
amplitude deviations. The detuned STIRAP 
rotations exhibit robustness over a larger 
range of deviations in the pulse parameters. 
The minimum error introduced by the finite length
detuned STIRAP can, in principle, be further reduced by 
increasing the single-photon detuning or employing 
techniques such as shortcuts to adiabaticity.

% \bibliographystyle{pra}
% \bibliography{ref}

%

\end{document}